\documentstyle[12pt,epsfig]{article}

\begin{document}
\renewcommand{\thefootnote}{*}

\newcommand{\lsim}{\raisebox{-0.13cm}{~\shortstack{$<$ \\[-0.07cm] $\sim$}}~}
\newcommand{\gsim}{\raisebox{-0.13cm}{~\shortstack{$>$ \\[-0.07cm] $\sim$}}~}
\newcommand{\nn}{\noindent}
\newcommand{\non}{\nonumber}
\newcommand{\ee}{e^+ e^-}
\newcommand{\ra}{\rightarrow}
\newcommand{\tb}{\tan \beta}
\newcommand{\s}{\smallskip}
\newcommand{\beq}{\begin{eqnarray}}
\newcommand{\eeq}{\end{eqnarray}}

\baselineskip=17pt

\begin{flushright}
CERN TH/2003--043\\
PM/02--03\\
March 2003\\
\end{flushright}

\vspace*{0.1cm}

\begin{center}

{\large\sc {\bf Higgs Physics at Future Colliders:}}

{\large\sc {\bf recent theoretical developments\footnote{Plenary talk given at 
the Conference ``Particles, Strings and Cosmology" (PASCOS), Bombay, India, 
3--8 January  2003.}}}

\vspace{0.5cm}

{\sc Abdelhak DJOUADI} 
\vspace*{2mm} 

Theory  Division, CERN, CH--1211 Geneva 23, Switzerland, \\
and\\
Laboratoire de Physique Math\'ematique et Th\'eorique, UMR5825--CNRS,\\
Universit\'e de Montpellier II, F--34095 Montpellier Cedex 5, France. 
\end{center} 

\vspace*{.3cm} 

\begin{abstract}
\nn I review the physics of the Higgs sector in the Standard Model and its
minimal  supersymmetric extension, the MSSM. I will discuss the prospects for
discovering the Higgs particles at the upgraded Tevatron, at the Large Hadron
Collider, and at a future high--energy $e^+e^-$ linear  collider with 
centre--of--mass energy in the 350--800 GeV range, as well as the possibilities
for studying their fundamental properties. Some emphasis will be put on the 
theoretical developments which occurred in the  last two years. 
\end{abstract}
\vspace*{.5cm}

\setcounter{footnote}{0}
\renewcommand{\thefootnote}{\arabic{footnote}}
\subsection*{1. A brief introduction}

The search for Higgs bosons is the primary mission of present and future
high--energy colliders. Detailed theoretical and experimental studies performed
in the last few years, have shown that the single neutral Higgs boson that is
predicted in the Standard Model (SM) \cite{HHG} could be discovered at the
upgraded Tevatron, if it is relatively light and if enough integrated luminosity
is collected \cite{Tevatron,Houches} and can be detected at the LHC
\cite{Houches,LHC} over its entire mass range 114.4 GeV $\lsim M_H \lsim 1$ TeV
in many redundant channels; see Fig.~1. In the context of the Minimal
Supersymmetric Standard Model (MSSM), where the  Higgs sector is extended to
contain two CP--even neutral Higgs bosons $h$ and $H$, a pseudoscalar $A$ boson
and a pair of charged scalar particles $H^\pm$ \cite{HHG}, it has been shown
that the lighter $h$ boson cannot escape detection at the LHC and that in large
areas of the parameter space, more than one Higgs particle can be found; see
Fig.~1.\s

\begin{figure}[hbtp]
\vspace*{-.6cm}
\begin{center}
\epsfig{file=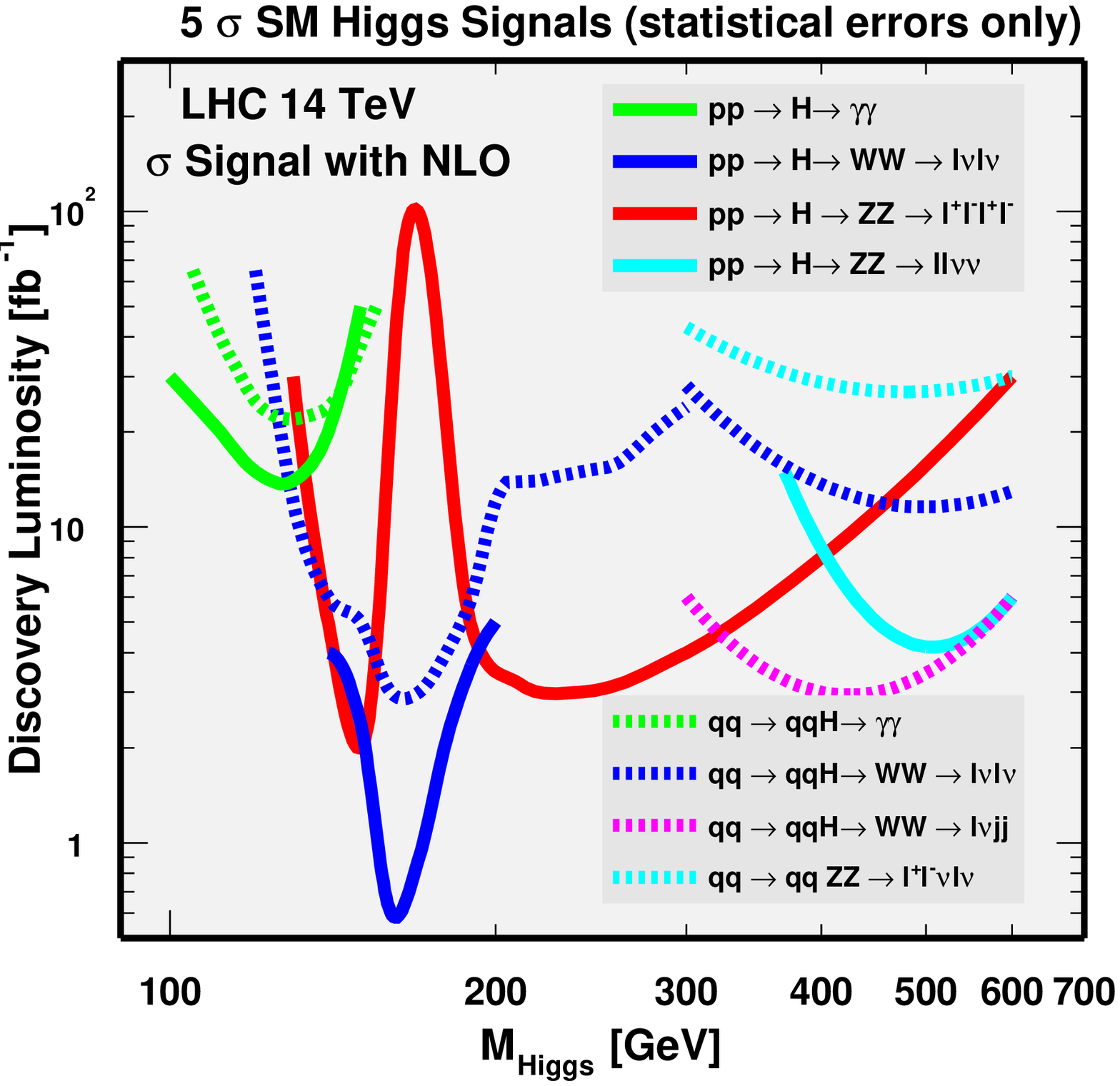,width=7.5cm,height=7.8cm}\hspace*{-3mm}
\epsfig{file=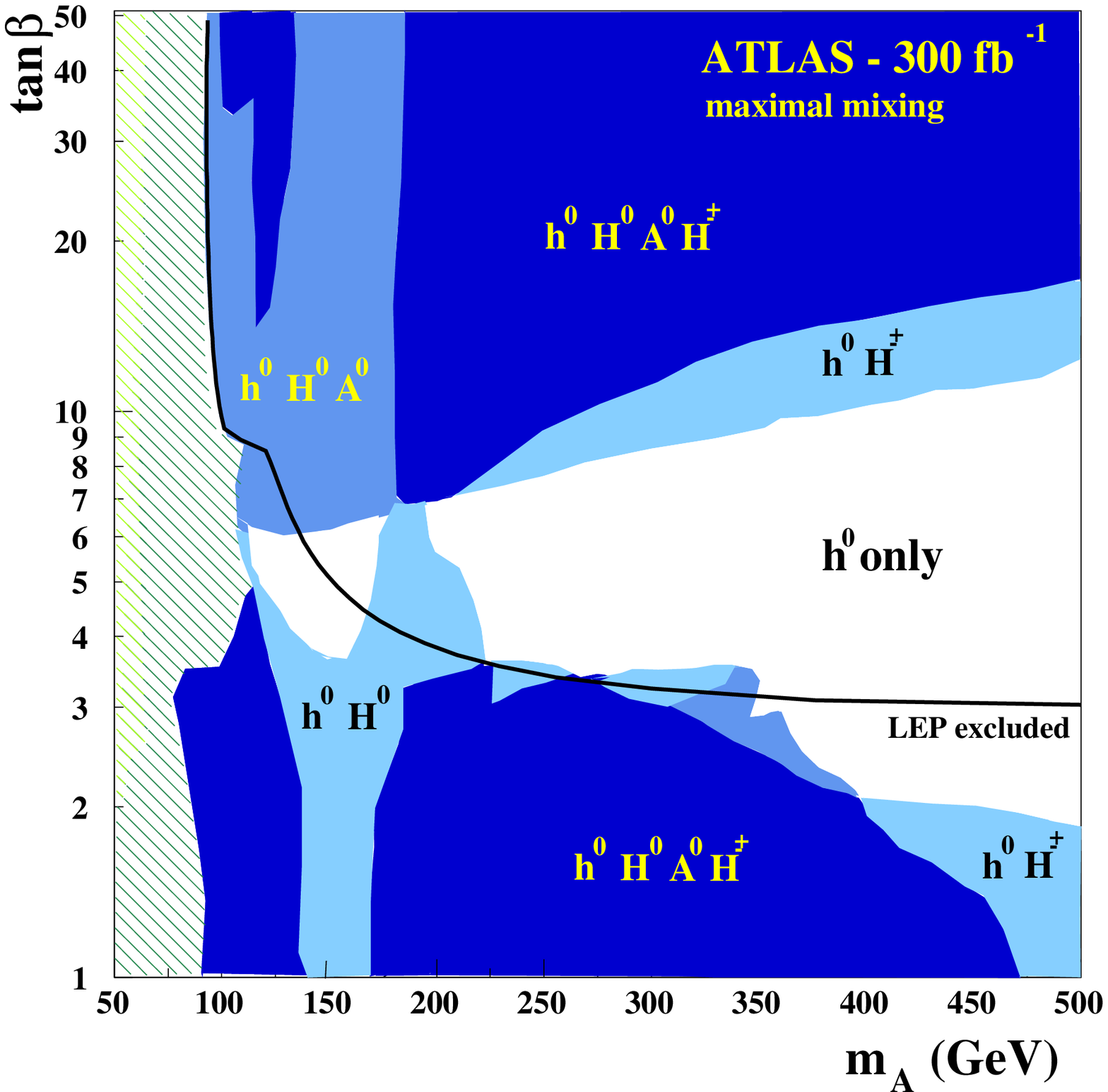,width=7.cm,height=7.1cm}
\end{center}
\vspace*{-7mm}
\caption{The integrated luminosity needed for the  discovery of the SM Higgs
boson  at the LHC in various production and decay channels (left) and the
number of  Higgs particles that can be detected in the MSSM  $(\tb,M_A)$ 
parameter space (right); from Ref.~\cite{LHCplots}.}
\vspace*{-3mm}
\end{figure}

Should we then declare that we have done our homework and wait peacefully for 
the LHC to start operation? Well, discovering the Higgs boson is not the entire
story, and another goal, just as important, would be to probe the electroweak
symmetry breaking mechanism in all its facets. Once the Higgs boson is found,
the next step would therefore be to perform very high precision measurements to
explore all its fundamental properties. To achieve this goal in great detail,
one needs to measure all possible cross sections and decay branching ratios of
the Higgs bosons to derive their masses, their total decay widths, their
couplings to the other particles and their self--couplings, their spin--parity
quantum numbers, etc.  This needs very precise theoretical predictions and
more involved theoretical and experimental studies. In particular, all possible
production and decay channels of the Higgs particles, not only the dominant and
widely studied ones allowing for clear discovery, should be investigated.  This
also requires complementary detailed studies at future $\ee$ linear colliders,
where the clean environment and the expected high luminosity allow for very
high precision measurements \cite{e+e-,Tesla}. 

In this talk, I will summarize the studies that were performed recently in the
SM and MSSM Higgs sectors\footnote{Other extensions have been discussed by Jack
Gunion \cite{Jack}, to whom we refer for details.}. In the next section, after
summarizing the present constraints, I will discuss the new  developments in
the calculation of the Higgs boson spectrum and decay branching ratios. In  in
sections 3 and 4, we will discuss the developments in Higgs  production at the
LHC and Tevatron hadron colliders and  at  a future $\ee$  machine with a c.m.
energy below 1 TeV. A brief conclusion will be presented in section 5.

\subsection*{2. Higgs spectrum and decay branching ratios}

In the SM, the profile of the Higgs boson is uniquely determined once $M_H$ is
fixed \cite{HHG}:  the decay width and branching ratios, as well as the
production cross sections, are given by the strength of the Yukawa couplings to
fermions and gauge bosons, which is set by the masses of these particles.  There
are two experimental constraints on this free parameter. 

The SM Higgs boson has been searched for at LEP in the Higgs--strahlung process
$\ee \to HZ$ for c.m.\, energies up to $\sqrt{s}=209$ GeV and with a
large collected luminosity. In summer 2002, the final results of the four LEP
collaborations were published [and some changes with respect to the original
publications occurred, in particular inclusion of more statistics, revision of
backgrounds, and reassessment of systematic errors]. When these results are
combined, an upper limit $M_H \geq 114.4$ GeV is established at the 95\%
confidence level \cite{Igo+LEPH}.  However, this upper limit, in the absence of 
additional events with respect to SM predictions, was expected to be $M_H>115.3$
GeV; the reason is that there is a $1.7\sigma$ excess [compared to the value 
$2.9\sigma$ reported at the end of 2000] of events for a Higgs boson mass in
the vicinity of $M_H=116$ GeV \cite{Igo+LEPH}.

The second constraint comes from the accuracy of the electroweak observables
measured at LEP, the SLC and the Tevatron, which provides sensitivity to $M_H$:
the Higgs boson contributes logarithmically, $\propto \log (M_H/M_W)$, to the
radiative corrections to the $W/Z$ boson propagators and alters these
observables\footnote{More details  are given in the talk of Guido Altarelli  at
this conference.}. The  status, as in summer 2002, is summarized in Fig.~2,
which shows the  $\Delta \chi^2$ of the fit to electroweak precision
measurements as a function of $M_H$ \cite{GA+LEPH}. When all available data
[i.e. the Z--boson pole LEP and SLC data, the measurement of the $W$ boson mass
and total width, the top--quark mass and the controversial NuTeV result] are
taken into account, one obtains a Higgs boson mass of $M_H=81^ {+42}_{-33}$
GeV, leading to a 95\% confidence level upper limit  of $M_H <193$ GeV. These
values are relatively stable when the NuTeV result is excluded from the fit, or
when a different value for the hadronic contribution to the QED coupling
$\alpha$ is used. 

However, theoretical constraints can also be derived from assumptions on the
energy range within which the SM is valid before perturbation theory breaks
down and New Physics should appear.  If $M_H \gsim 1$ TeV, the longitudinal $W$
and $Z$  bosons would interact strongly;  to ensure unitarity in their 
scattering at high  energies, one needs $M_H \lsim 710$ GeV at tree--level
\cite{unitarity}. In addition, the quartic Higgs self--coupling, which at the 
weak scale is fixed by $M_H$, grows logarithmically with energy and a cut--off
$\Lambda$ should be imposed before it grows beyond any bound.  The condition
$M_H \lsim \Lambda$ sets an upper limit at $M_H \sim 630$ GeV.   Furthermore,
top quark loops tend to drive the coupling to negative values, for which the
vacuum becomes unstable. Requiring the SM to be extended to the GUT scale,
$\Lambda \sim 10^{16}$ GeV, the Higgs mass should lie in the range
130 GeV $\lsim M_H \lsim 180$ GeV \cite{Trivial}. 

\begin{tabular}{ll}
\begin{minipage}{5.2cm}
\vspace*{-1.cm}
Figure 2:  The $\Delta \chi^2$ of the fit to electroweak  precision data as a
function of $M_H$. The  solid line is when all data are included and  the blue
band is the estimated theoretical  error from unknown higher order
corrections.  The effect of excluding the NuTeV measurement  and the use of a
different value for $\Delta  \alpha_{\rm had}$ are also shown. The vertical band
shows the 95\% CL exclusion  limit on $M_H$ from direct searches. From
Ref.~\cite{GA+LEPH}. 
\end{minipage}
&
\begin{minipage}{16cm}
\vspace*{-1cm}
\epsfig{file=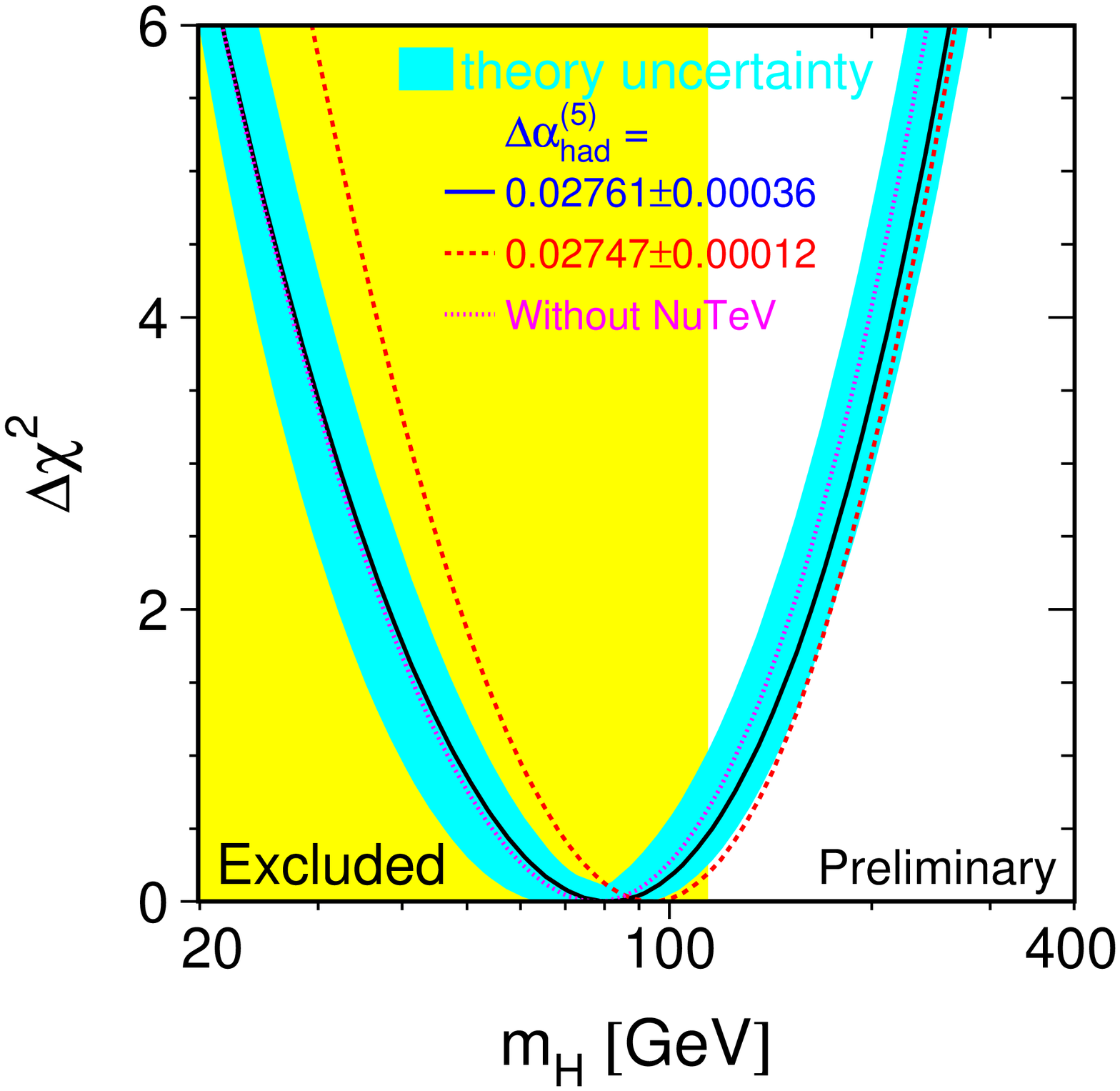,width=0.6\linewidth,height=10.cm} 
\end{minipage}
\vspace*{-2mm}
\end{tabular}

In the MSSM, two doublets of Higgs fields are needed to break the electroweak
symmetry, leading to two CP--even  neutral $h,H$ bosons, a pseudoscalar $A$
boson and a pair of charged  scalar particles, $H^\pm$ \cite{HHG}. Besides the
four masses, two additional  parameters define the properties of the particles:
a mixing angle $\alpha$ in the neutral CP--even sector, and the ratio of the two
vacuum expectation values, $\tb$. Because of supersymmetry constraints,  only
two of them, e.g.\, $M_A$ and $\tb$, are in fact independent at tree--level.
While the lightest Higgs mass is bounded by $M_h \leq M_Z$, the masses of the
$A,H$ and $H^\pm$ states are expected to be below ${\cal O}(1$ TeV). However,
mainly because of the heaviness of the top quark, radiative corrections are
very important: the leading part  grows as the fourth power of $m_t$ and
logarithmically with the common top squark mass $M_S$; the stop trilinear 
coupling $A_t$ also plays an important role and maximizes the correction for
the value $A_t \sim 2 M_S$. For a recent review, see Ref.~\cite{RCreview}. 

Recently, new calculations of the two--loop radiative corrections have been 
performed \cite{Pietro}. Besides the already known  ${\cal O} (\alpha_t
\alpha_s)$ correction, the contributions at ${\cal O}  (\alpha_t^2)$ and ${\cal
O} (\alpha_s \alpha_b)$ have been derived. By an  appropriate use of the
effective potential approach, one obtains simple analytic formulae for
arbitrary values of $M_A$ and of the parameters in the stop sector.  In a large
region of the parameter space, the ${\cal O} (\alpha_t^2)$ corrections are
sizeable, increasing the predicted value for $M_h$ [for given $\tb$ and $M_A$
inputs] by several GeV.  This is exemplified in Fig.~3, where the value of $M_h$
is shown as a function of $M_A$ for $\tb=2$ and 20 in various approximations.
As can be seen, the upper bound on $M_h$ can reach the level of 130 GeV if the
corrections due to  $\alpha_t$ are included. At large values of $\tb$ where the
Yukawa coupling of the $b$--quark becomes rather large, a further increase of a
few GeV is obtained if the  ${\cal O} (\alpha_s \alpha_b)$  correction is
included. 

\begin{tabular}{ll}
\begin{minipage}{5cm}
\vspace*{-.6cm}
Figure 3: The value of the lightest $h$ boson mass as a function of $M_A$ in
the MSSM for $\tb=2$ and~20 in the maximal mixing scenario $X_t \sim A_t \sim 2
M_S \sim 2$ TeV. The long--dashed line shows the result at one--loop, while the
full line shows the two--loop result including the full ${\cal O} (\alpha_t 
\alpha_s)$ and   ${\cal O}(\alpha_t^2)$ corrections; from Ref.~\cite{Pietro}.  
\end{minipage}
&
\begin{minipage}{16cm}
\vspace*{-.6cm}
\epsfig{figure=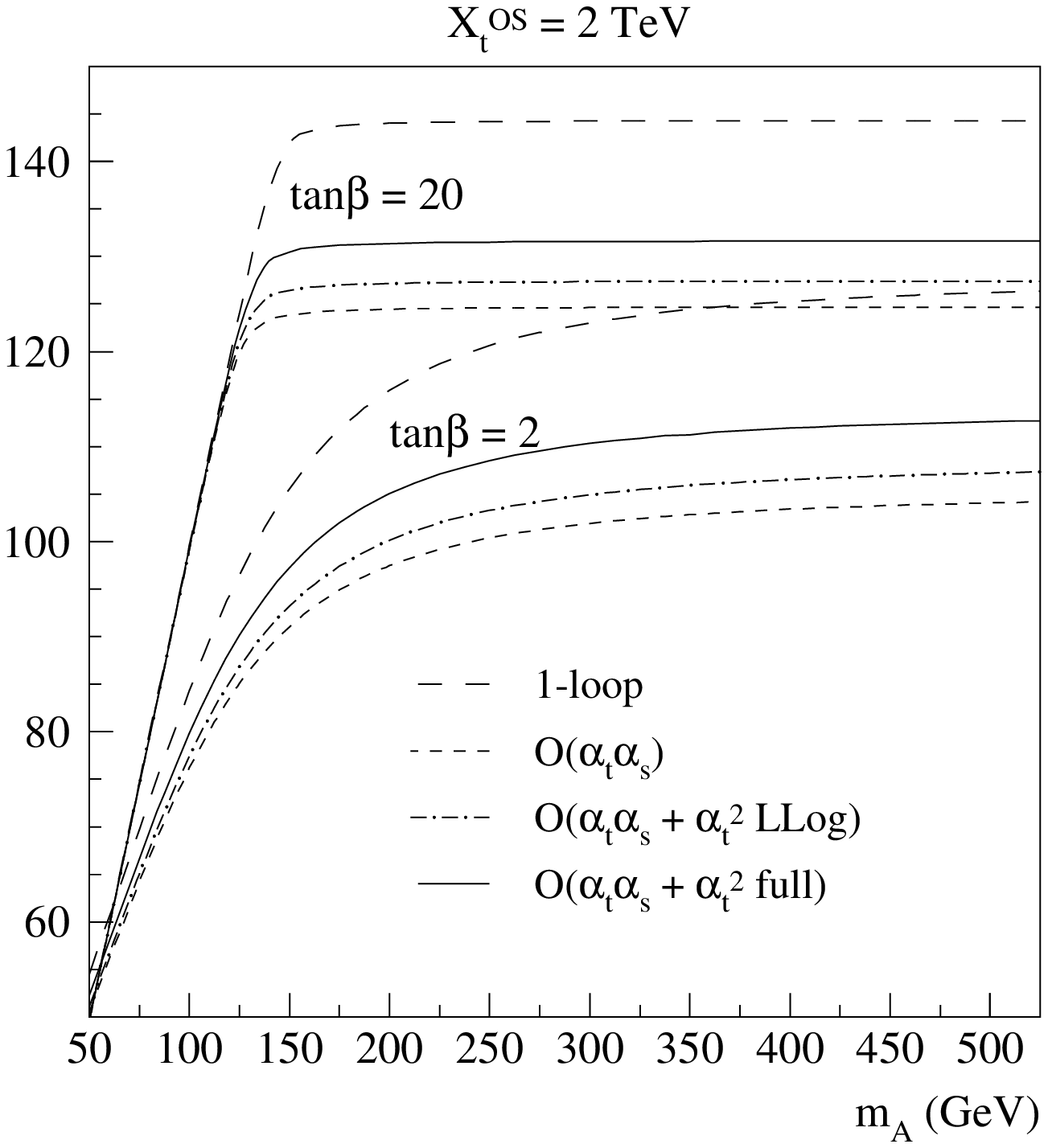,width=9cm,height=9cm}
\vspace*{-3mm}
\end{minipage}
\end{tabular}
\setcounter{figure}{3}

Note that these important radiative corrections are now being implemented in
the three new codes for the determination of the MSSM particle spectrum, which
appeared in the last year: {\tt Softsusy}, {\tt SuSpect} and {\tt Spheno}
\cite{codes}.  

The production and the decays of the MSSM Higgs bosons depend strongly on their
couplings to gauge bosons and fermions.  The pseudoscalar has no tree level
couplings to gauge bosons, and its couplings to down--(up)--type fermions are
(inversely) proportional to $\tb$. It is also the case for the couplings of the
charged Higgs particle to fermions, which are a mixture of scalar and
pseudoscalar currents and depend only on $\tb$. For the CP--even Higgs bosons,
the couplings to down--(up)--type fermions are enhanced (suppressed) with
respect to the SM Higgs couplings for $\tb >1$. They share the SM Higgs 
couplings to vector bosons since they are suppressed by $\sin (\beta-\alpha)$
and $\cos(\beta-\alpha)$ factors, respectively for $h$ and $H$.  If the
pseudoscalar mass is large, the $h$ boson mass reaches its upper limit [which
depends on the value of $\tb$] and the angle $\alpha$ reaches the value
$\alpha= \beta-\frac{1}{2}\pi$. The $h$  couplings to fermions and gauge bosons
are then SM--like, while the heavier CP--even $H$ and charged $H^\pm$ bosons 
become degenerate in mass with $A$. In this decoupling limit, it is very
difficult to distinguish the SM and MSSM Higgs  sectors. 

The constraints on the MSSM Higgs particles masses mainly come from the
negative searches at LEP2 \cite{Igo+LEPH} in the Higgs--strahlung process, $\ee
\to Z+h/H$, and pair production process, $\ee \to A+h/H$, with the Higgs bosons
mainly decaying into $b\bar{b}$ pairs [these processes will be discussed 
later]. In the decoupling limit where the $h$ boson has SM--like couplings to
$Z$ bosons, the limit $M_h \gsim 114.4$ GeV from the $\ee \to hZ$ process
holds. This constraint rules out $\tb$ values larger than $\tb \gsim 2.5$. From
the $\ee \to Ah$ process, one obtains the absolute limits $M_h \gsim 91$ GeV
and $M_A \gsim 91.9$ GeV, for a maximal $ZhA$ coupling. More details are given
in the talk by P. Igo-Kemenes. 

Let us now discuss the Higgs decay modes and branching ratios (BR)
\cite{decays} and start with the SM  case [Fig.~4]. In the ``low--mass" range,
$M_H \lsim 130$ GeV,  the Higgs boson decays into  a large variety of channels.
The main mode is by far the decay into  $b\bar{b}$ with BR\,$\sim$ 90\%
followed by the decays into $c\bar{c}$ and  $\tau^+\tau^-$ with BRs\,$\sim$
5\%. Also of significance is the top--loop  mediated decay into gluons, which
occurs at the level of $\sim$ 5\%.  The top and $W$--loop mediated
$\gamma\gamma$ and $Z \gamma$ decay modes are very rare with BRs of ${\cal O
}(10^{-3})$ [however, they lead to  clear signals and are interesting, since
they are sensitive to new heavy particles].  In the ``high--mass" range, $M_H
\gsim 130$ GeV, the Higgs bosons decay into $WW$ and $ZZ$ pairs, one of the
gauge bosons being possibly virtual  below the thresholds. Above the $ZZ$
threshold, the BRs are 2/3 for $WW$ and  1/3 for $ZZ$ decays, and the opening
of the $t\bar{t}$ channel for higher $M_H$ does not alter  this pattern
significantly.  In the low--mass range, the Higgs is very narrow, with
$\Gamma_H<10$ MeV, but this width becomes wider rapidly,   reaching 1 GeV at
the $ZZ$ threshold. For  very large masses, the Higgs  becomes obese, since
$\Gamma_H \sim M_H$, and can hardly be considered as a resonance.

In the MSSM [Fig.~5], the lightest $h$ boson will decay  mainly into fermion
pairs since $M_H \lsim$~130~GeV.  This is, in general, also the dominant decay
mode of the $A$ and $H$ bosons, since for $\tb \gg 1$,  they decay into $b
\bar{b}$ and $\tau^+ \tau^-$ pairs with BRs of the order of $\sim$ 90\% and
10\%, respectively. For large masses, the top decay channels $H, A \rightarrow
t\bar{t}$ open up, yet they are suppressed  for large $\tb$. [The $H$ boson can
decay into gauge bosons or $h$ boson pairs, and the $A$ particle into $hZ$
final states; however, these decays are strongly suppressed for $\tb \gsim
3$--$5$ as is suggested by LEP2.]  The $H^\pm$ particles decay into fermions
pairs: mainly $t\bar{b}$ and $\tau \nu_{\tau}$ final states for $H^\pm$ masses,
respectively, above and below the $tb$ threshold.  [If allowed kinematically,
they can also decay  into $hW$ final states for $\tb \lsim 5$.] Adding up the
various decays, the widths of all five Higgsses remain  rather narrow  [very
small for $h$ and a few tens of GeV for $H,A$ and $H^\pm$ masses of  ${\cal
O}(1$ TeV)].

Other possible decay channels for the heavy $H, A$ and $H^\pm$ states, are
decays into light charginos and neutralinos, which could be important if  not
dominant \cite{SUSYdecays}; decays of the $h$ boson into the lightest
neutralinos (LSP) can also be important, exceeding 50\% in some parts of the 
parameter space and altering the searches at hadron  colliders as will be
discussed later. SUSY particles can also affect the BRs of the loop--induced
modes \cite{SUSYloops}.  

The various decay widths and branching ratios of the SM and MSSM Higgs boson
can be calculated in a very precise way with the Fortran code {\tt HDECAY}
\cite{HDECAY}, in which all relevant processes are implemented [including
many-body and SUSY channels] and all important higher order [QCD and Higgs]
effects.  Recently, it has been upgraded to include new features [in addition
to a reorganization of the program to make it more user--friendly] such as a
more precise determination of the MSSM Higgs spectrum, the SUSY--QCD
corrections to decays in $b\bar{b}$, and decays in the gauge--mediated SUSY
breaking scenario.  

\begin{figure}[htbp]
\begin{center}
\vspace*{-.5cm}
\epsfig{file=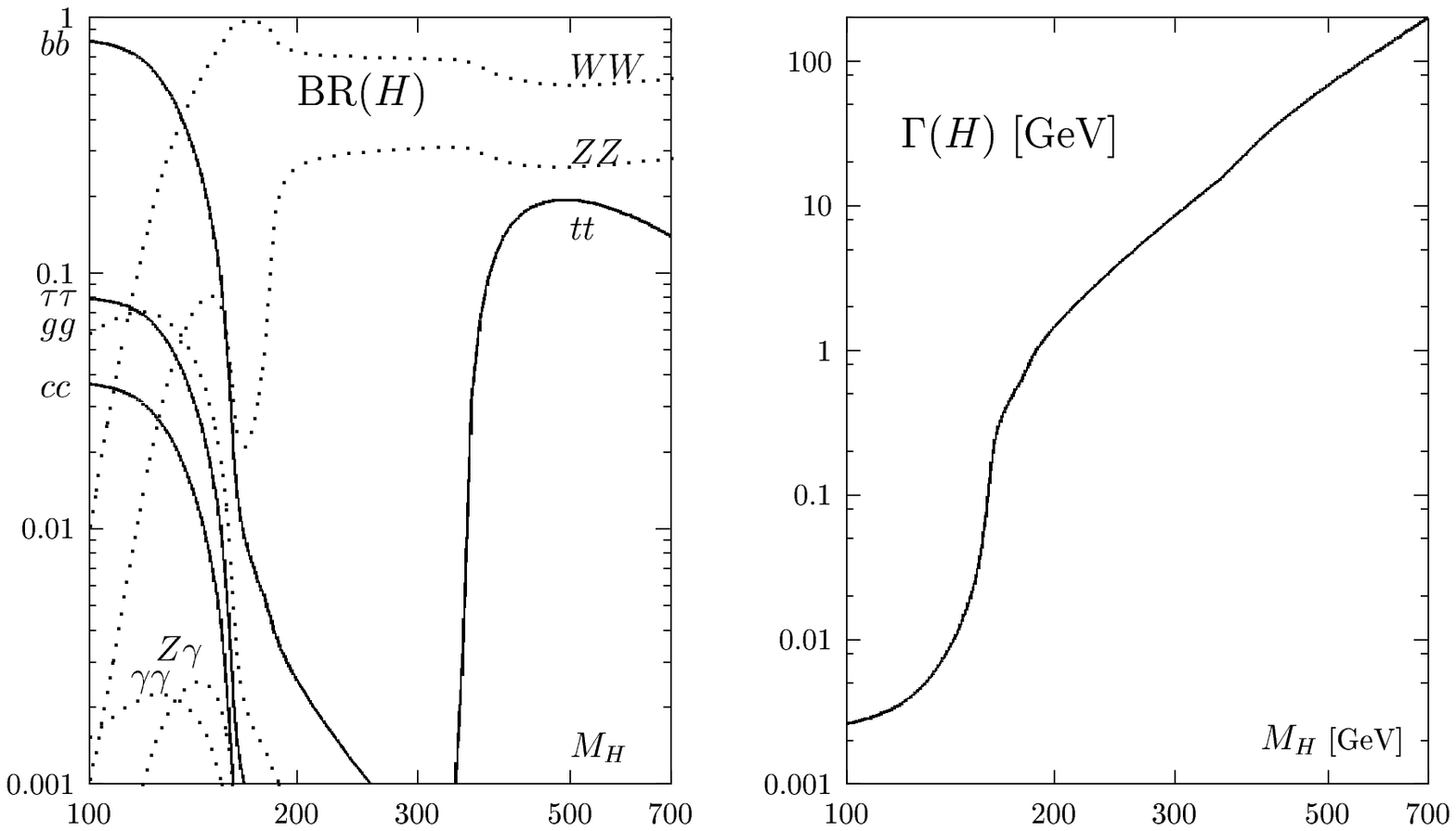,width=15cm,height=7cm}
\end{center}
\vspace*{-7mm}
\caption[]{The decay branching ratios (left) and the total decay width (right) 
of the SM Higgs boson as a function of its mass.}
\label{fig:hbrth}
\vspace*{-.3cm}
\end{figure}

\begin{figure}[htbp!]
\begin{center}
\vspace*{-2.9cm}
\hspace*{-2cm}
\epsfig{file=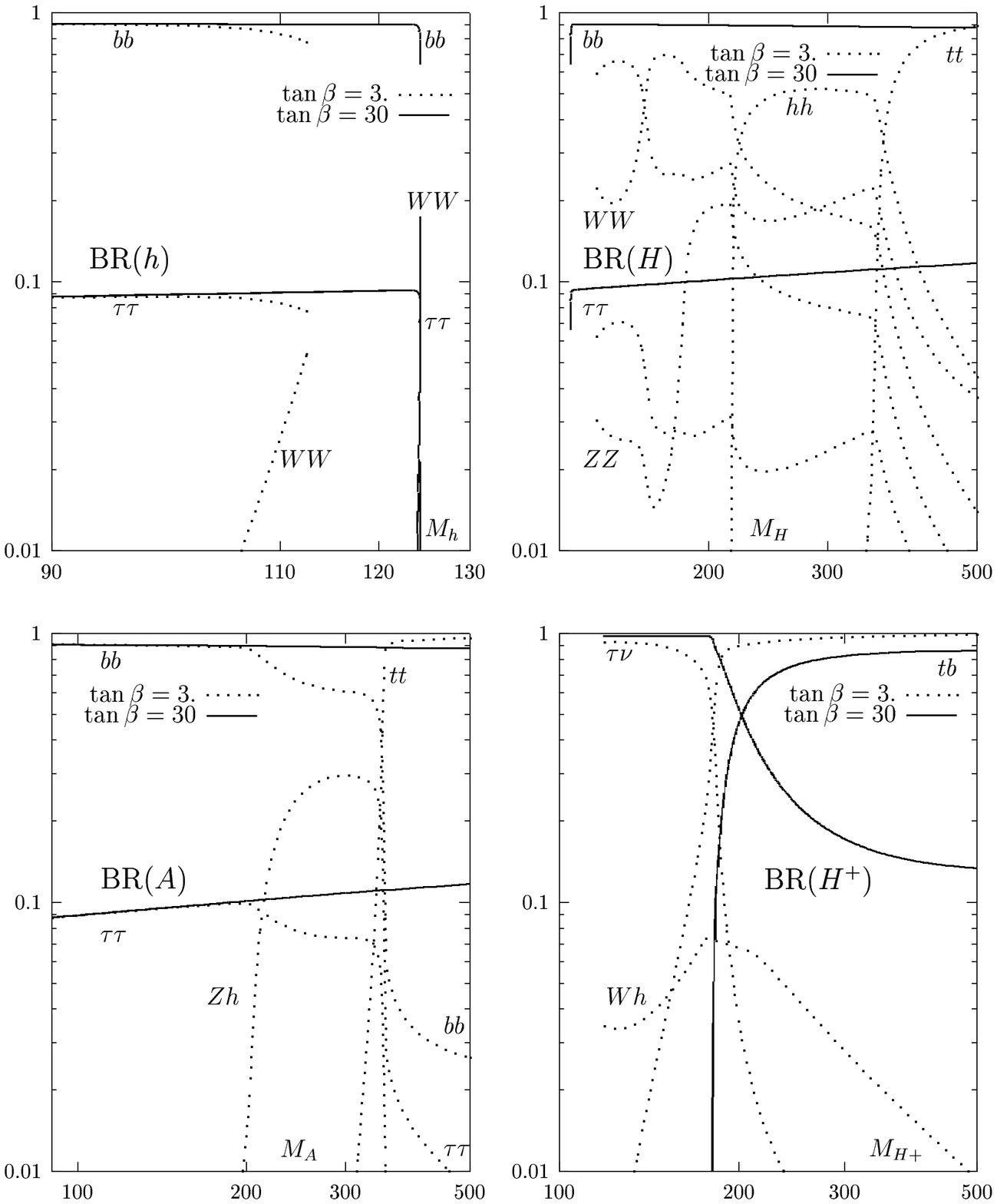,width=16.cm,height=18cm} 
\end{center}
\vspace*{-4.1cm}
\caption{Dominant MSSM Higgs bosons decay branching ratios as functions of 
the Higgs boson masses for $\tb=3$ and 30.}
\label{fig:mssmbr}
\label{fig:MSSMdecays} 
\vspace*{-.1cm}
\end{figure}

\subsection*{3. Higgs production and measurements at hadron colliders} 

The production mechanisms for the SM Higgs bosons  at hadron colliders are 
\cite{P1}:
\vspace*{-3mm}
\begin{eqnarray}
\begin{array}{lccl}
(a) & \ \ {\rm gluon~gluon~fusion} & \ \ gg  \ \ \ra & H \nonumber \\
(b) & \ \ {\rm association~with}~W/Z & \ \ q\bar{q} \ \ \ra & V + H \nonumber\\
(c) & \ \ WW/ZZ~{\rm fusion}       & \ \ VV \  \ra &  H \nonumber \\
(d) & \ \ {\rm association~with~}Q\bar{Q} & gg,q\bar{q}\ra & Q\bar{Q}+H
\nonumber
\end{array}
\end{eqnarray}
\vspace*{-4mm}

\nn The cross sections are shown in Fig.~6 for the LHC with $\sqrt{s}=14$ TeV 
and for the Tevatron with $\sqrt{s}=2$ TeV as functions of the Higgs boson 
masses; from Ref.~\cite{Pplot}. \ \ \ \ \ \ \ \ \ \ \ \ \ \ \ \ \ \ \vspace*{-.7cm}
\begin{figure}[htbp]
\hspace*{-1.5cm}
\epsfig{file=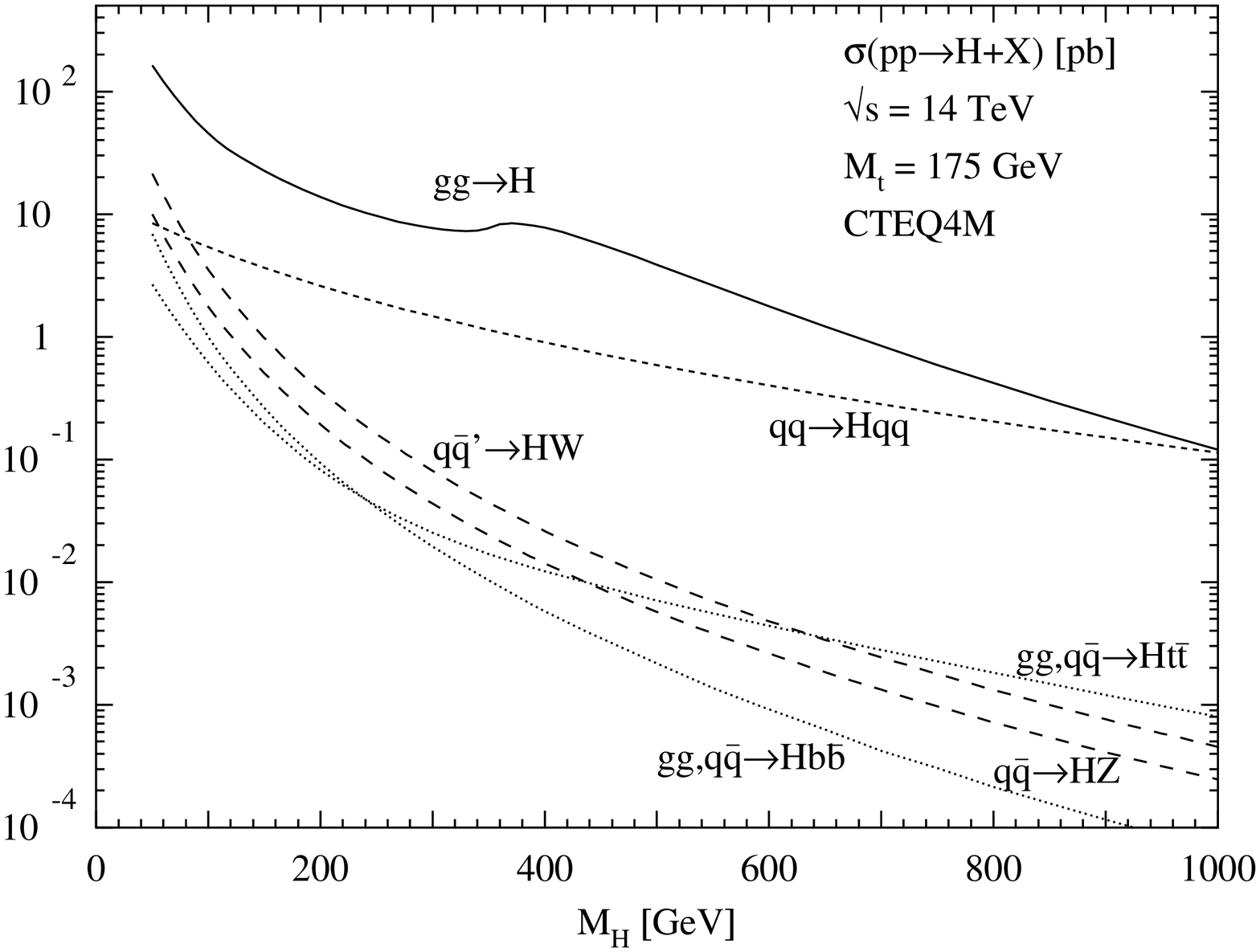,bbllx=3,bblly=3,bburx=410,bbury=420,width=5.cm,angle=-90,clip=}\hspace*{2.5cm}
\epsfig{file=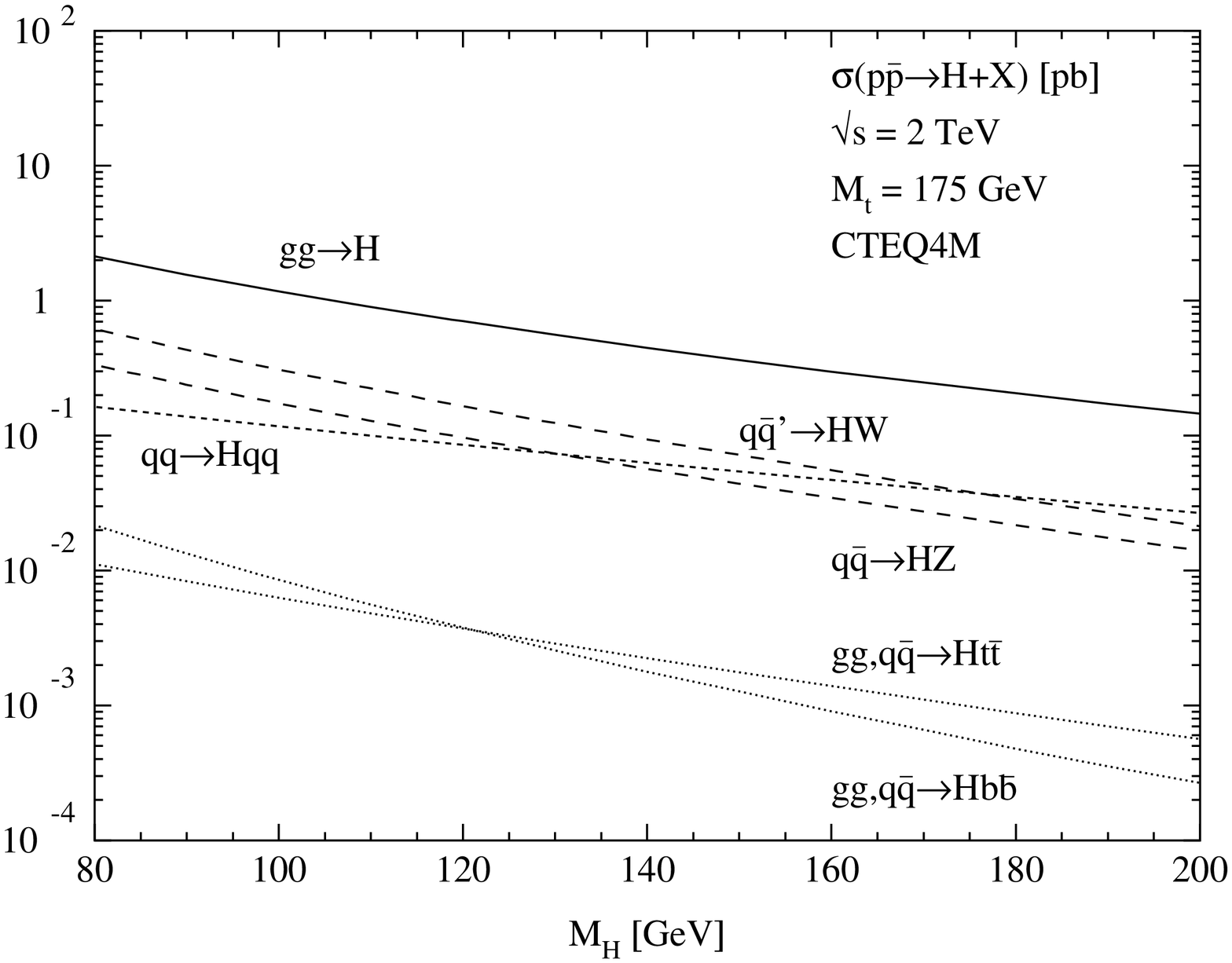,bbllx=3,bblly=3,bburx=410,bbury=420,width=5.cm,angle=-90,clip=} 
\vspace*{1.1cm}\\[.1cm]
\caption{SM Higgs production cross sections at the LHC 
(left) and the Tevatron (right).}
\vspace*{-3mm}
\end{figure}
Let us discuss the main features of each channel and highlight the new
developments:
 
{\bf a)} At the LHC, the dominant production process, up to masses $M_H \lsim 
700$ GeV, is by far the $gg$ fusion  mechanism. The most promising signals are $H
\to \gamma \gamma$ in the mass range below 130 GeV; for larger masses it
is $H \ra Z Z^{(*)} \ra 4 \ell^\pm$, with $\ell=e,\mu$, which from $M_H \gsim
500$ GeV can be complemented by $H \to ZZ \to \nu\bar{\nu} \ell^+ \ell^-$ and 
$H \to WW \to \nu \ell jj$. The QCD next--to--leading order (NLO) corrections
should  be taken into account since they lead to an increase of the  cross
sections by a factor of $\sim 1.7$ \cite{P2}. Recently, the three--loop
corrections have been calculated [a real ``tour de force"] in the heavy--top
limit and shown to increase the rate by an additional 30\% \cite{P3}. This 
results in a nice convergence of the perturbative series and a strong
reduction  of the scale uncertainty, which is the measure of higher order
effects; see  Fig.~7. The corrections to the differential  distributions have
also been recently calculated at NLO and shown to be rather large \cite{P4}. 

{\bf b)} The associated production with gauge bosons, with $H \to b\bar{b}$ 
[and possibly $H \to WW^* \to \ell^+ \nu jj$], is the most relevant mechanism
at the Tevatron \cite{Tevatron}, since the  dominant $gg$ mechanism with the
same final state has too large a QCD  background. The QCD corrections, which
can be inferred from Drell--Yan production, are at the level of 30\%
\cite{QCDpp}.   At the LHC, this process plays only a marginal role; however,
it could be  useful in the MSSM, if the Higgs decays invisibly into the LSPs,
as recently shown \cite{invisible}. 

\nn
\begin{tabular}{ll}
\begin{minipage}{5.5cm}
Figure 7: SM Higgs boson production cross sections in the $gg$ fusion process 
at the LHC as a function  of $M_H$: LO (dotted), NLO (dashed) and NLLO (full).
The upper (lower) curves  are for the choice of the renormalization and
factorization scales $\mu=\frac{1} {2} M_H$ ($2M_H$).
From Harlander and Kilgore in Ref.~\cite{P3}. 
\end{minipage}
&
\begin{minipage}{10cm}
\epsfig{file=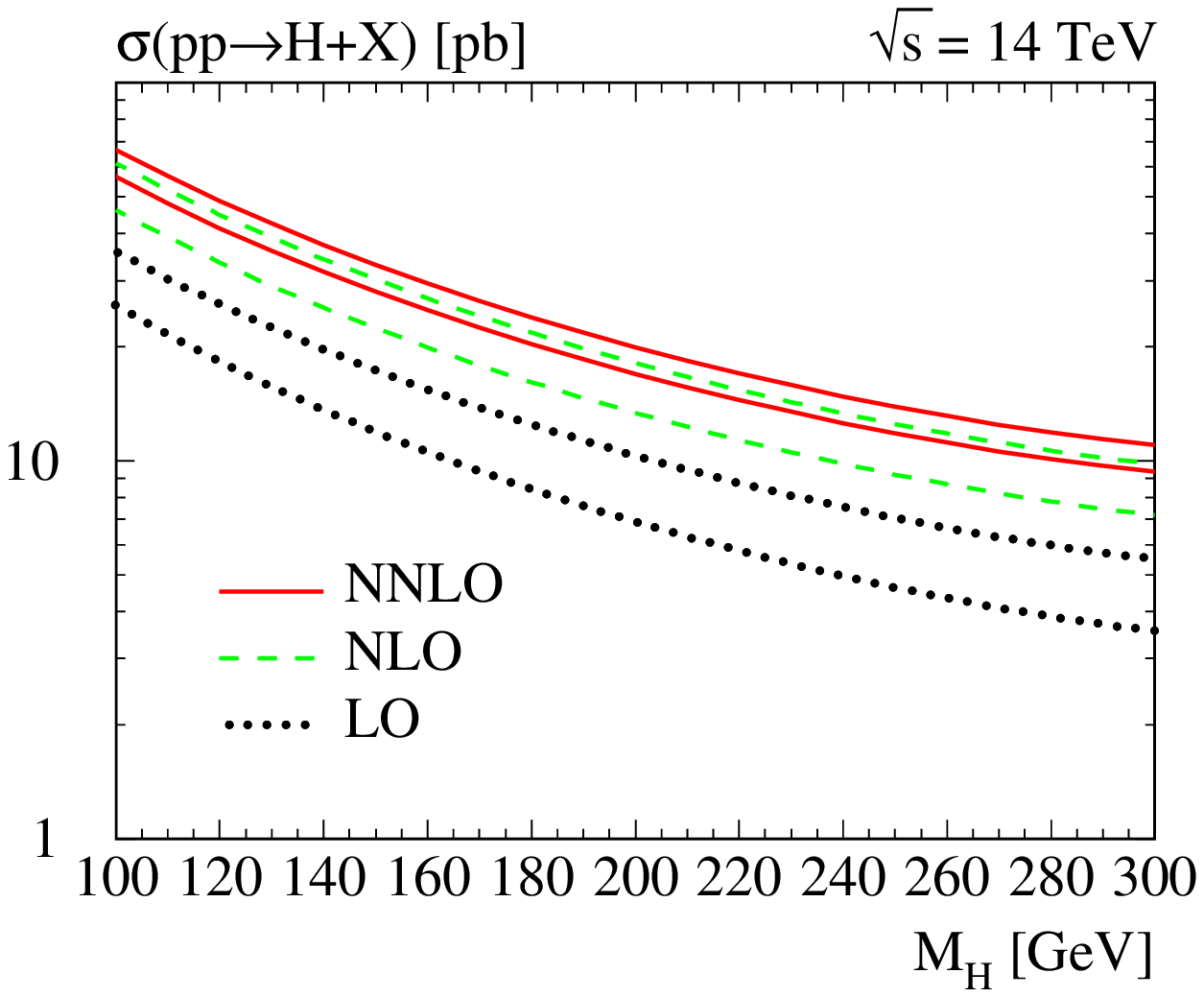,bbllx=110,bblly=265,bburx=465,bbury=560,width=9.5cm} 
\end{minipage}
\vspace*{3mm}    
\end{tabular}

{\bf c)} The $WW/ZZ$ fusion mechanism has the second largest cross section at
the LHC. The QCD corrections, which can be obtained in the structure--function
approach, are at the level of 10\% and thus small \cite{QCDpp}. For several 
reasons, the interest in
this process has grown in recent years: it has a large
enough cross section [a few picobarns for $M_H \lsim 250$ GeV], rather small
backgrounds [comparable to the signal] allowing precision measurements, one can
use forward--jet tagging of mini--jet veto for low luminosity, and one can
trigger on the central Higgs decay products \cite{WWfusion1}. In the past, it
has been shown that the decay $H \to \tau^+ \tau^-$ and possibly $H \to \gamma
\gamma , ZZ^*$ can be detected and could allow for coupling measurements
\cite{Houches,Dieter}. In the last two years, several ``theoretical" analyses have
shown that various  other channels can also be detected in some cases
\cite{WWfusion}:  $H \to WW^*$ for $M_H \sim$ 125--180 GeV, $H \to \mu^+ \mu^-$
[for  second--generation coupling measurements], $H \to b\bar{b}$ [for the 
$b\bar{b}H$ Yukawa coupling] and $H \to $ invisible [if forward--jet trigger]. 
However, more detailed analyses, in particular experimental  simulations [some 
of which have started \cite{Karl} already] are necessary to assess more firmly 
the potential of this channel.  

{\bf d)} Finally, Higgs boson production in association with top quarks, with
$H \to \gamma \gamma$ or $b\bar{b}$, can be observed at the LHC and would allow
the measurement of the important top Yukawa coupling. The cross  section is
rather involved at tree--level since it is a three--body process, and the
calculation of the NLO corrections was a real challenge, since one had to deal
with one--loop corrections involving pentagonal diagrams and real corrections
involving four particles in the final state. This challenge was taken up by two
groups [of US ladies and DESY gentlemen], and this calculation was completed
last year \cite{ttH}. The $K$--factors turned out to be rather small, $K\sim
1.2$ at the LHC and $\sim 0.8$ at the Tevatron [an example that  $K$--factors can also
be less than unity]. However, the scale dependence is drastically reduced from
a factor 3 at LO to the level of 10--20\% at NLO; see Fig.~8. 

\nn
\begin{tabular}{ll}
\begin{minipage}{5.5cm}
Figure 8: SM--Higgs boson production cross sections in  the $t\bar{t}H$ 
process at the LHC as a function  of the renormalization/factorization scale
$\mu$. The full (dashed) curves are for the NLO (LO) rates. From S. Dawson et
al. in Ref.~\cite{ttH}.
\end{minipage}
&
\begin{minipage}{10cm}
\includegraphics[width=20pc]{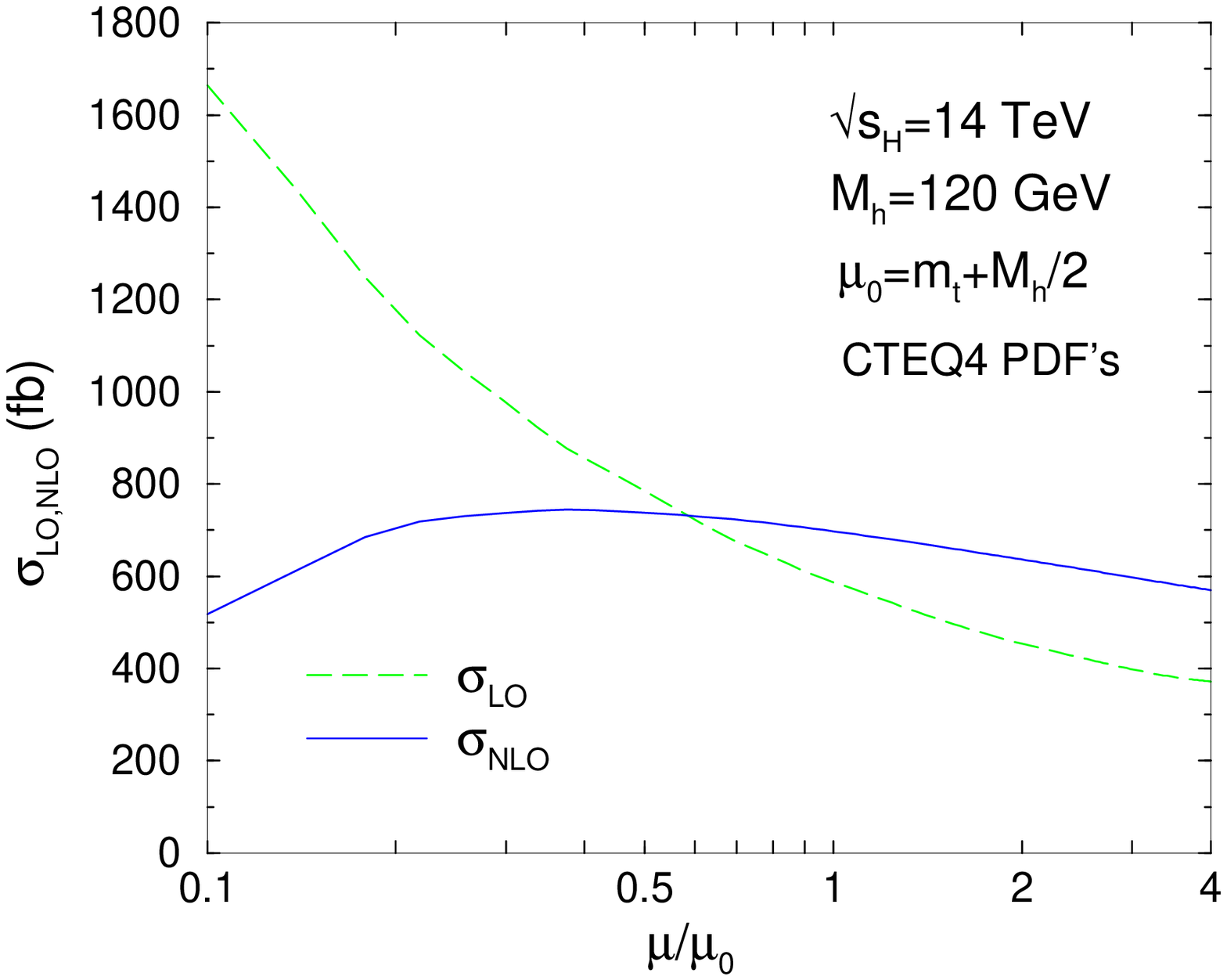}
\vspace*{.2cm}
\end{minipage}
\end{tabular}

Let us now turn to the measurements that can be performed at the LHC. We will
mostly rely on the analysis of Ref.~\cite{Dieter} and assume a large, ${\cal
O} (200)$ fb$^{-1}$, luminosity. 

$\bullet$ The Higgs boson mass can be measured with a very good accuracy. For
$M_H \lsim 400$ GeV, where $\Gamma_H$  is not too  large, a precision of
$\Delta M_H/M_H \sim 0.1$\% can be achieved in $H \to ZZ^{(*)} \to 4\ell^\pm$. 
In the ``low--mass" range, a slight improvement can be obtained by  considering
$H \to \gamma \gamma$. For $M_H \gsim 400$ GeV, the precision starts to
deteriorate because of the smaller rates. However, a precision of the order of
1\% can still be obtained up to $M_H\sim 800$ GeV if theoretical errors, such
as width effects, are not taken into account.  

$\bullet$ Using the same process, $H \to ZZ^{(*)} \to 4\ell^\pm$, the total
Higgs width can be measured for masses above $M_H \gsim 200$ GeV, when it is
large enough. While the precision is very poor near this mass value [a factor 
of 2], it improves to reach the level of $\sim 5$\% around $M_H \sim 400$ 
GeV. Here again, the theoretical errors are not taken into account.  

$\bullet$ The Higgs boson spin can be measured by looking at angular
correlations between the fermions in the final states in $H \to VV \to 4f$ 
\cite{spin}.  However the cross sections are rather small and the
environment too difficult.  Only the measurement of the decay planes of the two
$Z$ bosons decaying into four leptons seems promising. 

$\bullet$ The direct measurement of the Higgs couplings to gauge bosons and
fermions is possible, but with rather poor accuracy. This is due to the limited
statistics, the large backgrounds, and the theoretical uncertainties from the
limited precision on the parton densities and the higher--order radiative
corrections. An example of determination of cross sections times branching
fractions in various channels at the LHC is shown in Fig.~9. [Note that
experimental analyses accounting for the backgrounds and for the detector
efficiencies, as well as further theoretical studies for the signal and
backgrounds, need to be performed to confirm these values.] To reduce some
uncertainties,  it is more interesting to measure ratios of cross sections where
the   normalization cancels out. One can then make, in some cases, a
measurement  of ratios of BRs at the level of  10\%.

\nn
\begin{tabular}{ll}
\begin{minipage}{5.cm}
\vspace*{-5mm}
Figure 9: Expected relative errors on the determination of  $\sigma \times
{\rm BR}$ for various Higgs search channels at the LHC  with 200 fb$^{-1}$
data. Solid lines are for $gg$ fusion, dotted lines are for $t\bar{t}H$
associated production with $H \to b\bar{b}$ and $WW$, and dashed lines are the
expectations for the weak boson fusion process; from Ref.\, \cite{Dieter}. 
\end{minipage}
&
\begin{minipage}{16cm}
\includegraphics[width=7.6cm, angle=90]{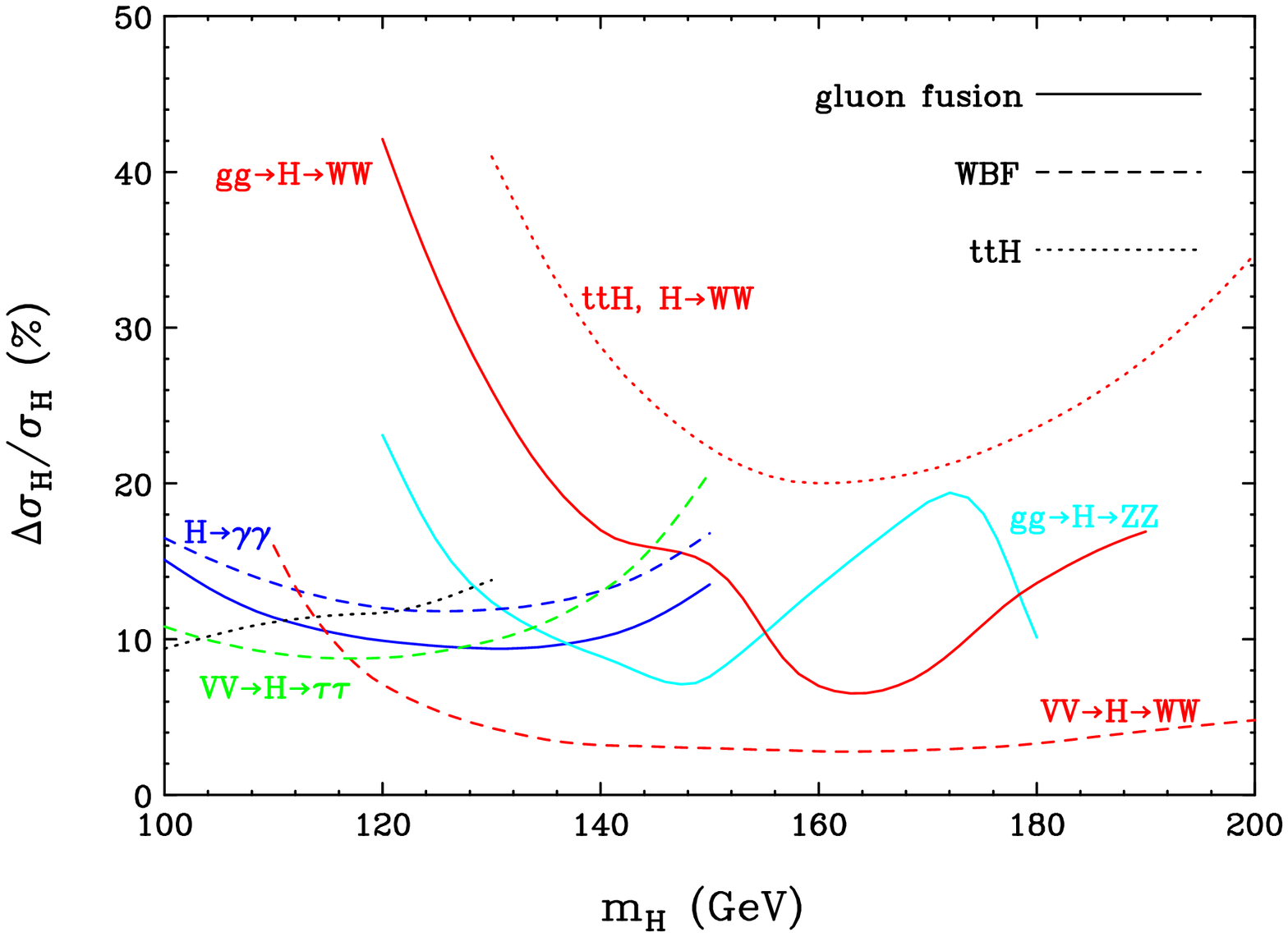}
\vspace*{.3cm}
\end{minipage}
\end{tabular} 

What happens in the case of the MSSM? The production processes for the $h,H$
bosons are practically the same as for the SM Higgs. However, for large $\tb$
values, one has to take the $b$ quark [whose couplings are strongly enhanced]
into account: its loop contributions in the $gg$ fusion process [and also the
extra contributions from squarks loops, which however decouple for high squark
masses] and associated production with $b\bar{b}$ pairs. The cross sections for
the associated production with $t\bar{t}$ pairs and $W/Z$ bosons as well as the
$WW/ZZ$ fusion processes, are suppressed for at least one of the particles as a
result of the coupling reduction.  Because of CP invariance, the $A$ boson can
be produced only in the $gg$ fusion and in association with heavy quarks
[associated production with a CP--even Higgs particle, $pp \to A+h/H$, is also
possible but the cross section is too small]. For high enough $\tb$ values and
for $M_A \gsim (\lsim) 130$ GeV, the $gg/q\bar{q}\ra b\bar{b}+A/H(h)$ and $gg
\to A/H(h)$ processes become the dominant production mechanisms.  The $H^\pm$
bosons are accessible in top decays, $t \ra H^+b$, if they are light enough,
otherwise they can be produced directly in the [properly combined] processes
$gb \to tH^-$ or $qq/ gg \to H^-t \bar{b}$. 

The various detection signals at the LHC are as follows \cite{Houches,LHC}. Since the
lightest Higgs boson mass is always smaller than $\sim 130 $ GeV, the $WW$ and
$ZZ$ signals cannot be used. Furthermore, the $hWW (h\bar{b}b)$ coupling is
suppressed (enhanced), leading to a smaller $\gamma \gamma$ branching ratio than
in the SM, which makes the search in this channel more difficult. If $M_h$ is close
to its maximum value, $h$ has SM--like couplings and the situation is similar to
the SM case with $M_H \sim $ 100--130 GeV.   For the $A$ and $H$ boson, since
their couplings to gauge bosons and are either absent or suppressed, the
gold--plated $ZZ$ signal is lost. In addition,  BR($A/ H \ra \gamma \gamma$)
are suppressed and these signals cannot be used. One then has to rely on the
$A/H  \ra \tau^+ \tau^-$ or even $\mu^+ \mu^-$ channels for large $\tb$
values.  [The decays $H \ra hh \ra b \bar{b}b\bar{b}$, $A \ra hZ \ra Zb
\bar{b}$ and $H/A \ra t\bar{t}$ have rates too small, in view of the LEP2
constraints]. Light $H^\pm$ particles can be observed \cite{P5} in the
decays $t \ra H^+b$ with $H^-\ra \tau \nu_\tau$, and heavier ones can be probed
for $\tb \gg 1$, by  considering $gb \to t H^-$ and $gg \ra t \bar{b}
H^-$  with $H^-\ra \tau \nu_\tau$ [using $\tau$ polarization] or 
$\bar{t}b$. A summary is given in Fig.~10.

Note that, in the situation where the pseudoscalar Higgs mass is small, $M_A 
\lsim 150$ GeV, and $\tb$ is large, $\tb \gsim 10$--30, all Higgs bosons will have
masses in the 100--150 GeV range and would couple strongly [in an almost
complementary way] into gauge bosons and third--generation fermions. In this
``intense coupling regime", all Higgs particles will be produced in many
competitive channels and a signal for one process can act as a background for
the other.  In addition, the $\gamma \gamma$ [and $ZZ^*$] BRs for all neutral
Higgs  particles can be suppressed at the same time and the decay widths of the
states  can be relatively large; this makes the searches slightly more involved
\cite{intense}. 

The whole previous discussion assumes that Higgs decays into SUSY  particles
are kinematically inaccessible. This seems to be  unlikely since at  least the
decays of the heavier $H,A$ and $H^\pm$ particles into charginos  and
neutralinos should be possible \cite{SUSYdecays}. Preliminary analyses show
that  decays into neutralino/chargino final states $H/A \to \chi_2^0 \chi_2^0
\to 4  \ell^\pm X$ and $H^\pm \to \chi_2^0 \chi_1^\pm \to 3 \ell^\pm X $ can
be  detected in some cases. It is also possible that the lighter $h$ decays
invisibly into the lightest neutralinos or sneutrinos.   If this scenario is
realized, the discovery of these Higgs particles will be more challenging.
Light SUSY particles can also alter the loop--induced production and decay
rates. For instance, light top squarks can couple strongly to the $h$ boson,
leading to a possibly strong suppression of the product $\sigma( gg \to
h)\times {\rm BR} (h \to \gamma \gamma)$ compared to the SM case
\cite{SUSYloops}. 

MSSM Higgs boson detection from the cascade decays of strongly interacting 
supersymmetric particles, which have large production production rates at the
LHC, is also possible.  In particular, the lighter $h$ boson and the heavier
$A,H$ and $H^\pm$ particles with $M_\Phi \lsim 200$ GeV, can be produced from
the decays of squarks and gluinos into the heavier charginos/neutralinos, which
then decay into the  lighter ones and Higgs bosons. This can occur either in
``little cascades" $\chi_2^0, \chi_1^\pm \to \chi_1^0 + \Phi$, or in 
``big cascades"  $\chi_{3,4}^0, \chi_2^\pm \to \chi_{1,2}^0, \chi_1^\pm +
\Phi$.  Recent studies \cite{cascade} show  that these processes can be
complementary to the direct production ones  in some areas of the MSSM
parameter space [in particular one can probe the region $M_A \sim 150$ GeV
and $\tb \sim 5$]; see Fig.~10.

Finally, at the Tevatron Run II, the search for the CP--even $h$ and $H$ bosons
in the MSSM will be more difficult than in the SM, because of the reduced
couplings to gauge bosons, unless one of the Higgs particles is SM--like.
However, associated production with $b\bar{b}$ pairs, $pp \to b\bar{b}+A/h\,
(H)$ in the low (high) $M_A$ range, with the Higgs bosons decaying into
$b\bar{b}$ pairs, might lead to a visible signal for rather large $\tb$ values
and $M_A$ values below the 200 GeV range.  The $H^\pm$ boson would also be
accessible in top--quark decays, for large or small values of $\tb$, for which
the  BR$(t \to H^+b)$ is large enough.

\nn 
\begin{tabular}{ll}
\begin{minipage}{6cm}
{\mbox Figure 10:~The~areas~in} the $(M_A, \tb)$ parameter space where the MSSM Higgs
bosons can be discovered at CMS with an integrated luminosity of 100 fb$^{-1}$.
Various detection channels are shown in the case of the standard searches. The 
right--hatched and cross--hatched regions show the areas where only the 
lightest $h$ boson can be observed in these channels. The left--hatched area 
is the region where the $H,A$ can be observed through the (big) cascade decays 
of squarks and gluinos in  some MSSM scenario.
\end{minipage}
&
\begin{minipage}{16cm}
\epsfig{file=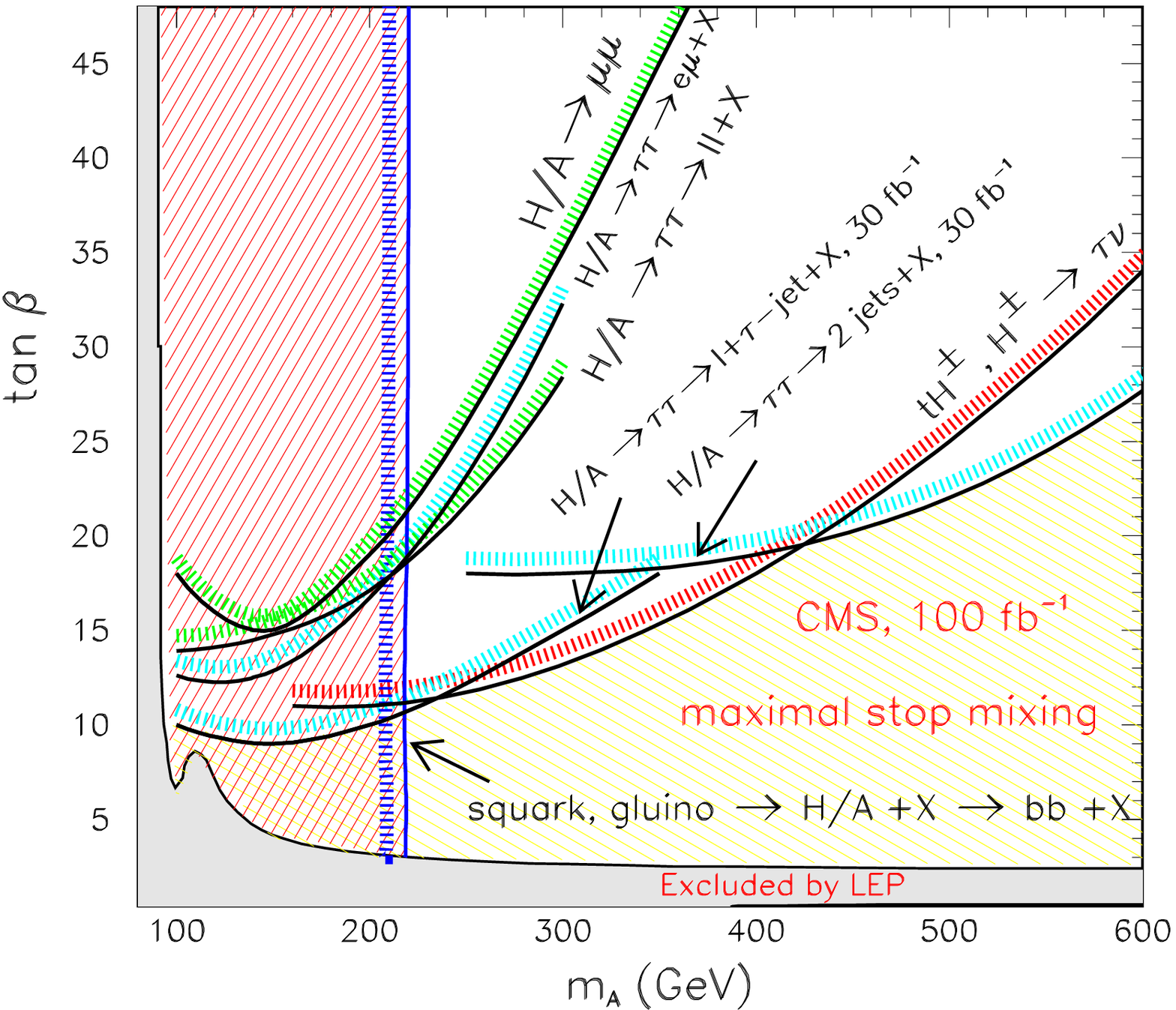,width=10cm}
\end{minipage}
\vspace*{5mm}
\end{tabular}

\subsection*{4. Higgs production at $e^+ e^-$ colliders}

At $\ee$ linear colliders operating in the 300--800 GeV energy range,  the main
production mechanisms for SM--like Higgs particles are \cite{E1}
\begin{eqnarray} 
\begin{array}{lccl} 
(a)  & \ \ {\rm bremsstrahlung \ process} & \ \ \ee & \ra (Z) \ra Z+H \non \\ 
(b)  & \ \ WW \ {\rm fusion \ process} & \ \ \ee & \ra \bar{\nu} \ \nu \ (WW) 
\ra \bar{\nu} \ \nu \ + H \non \\ 
(c)  & \ \ ZZ \ {\rm fusion \ process} & \ \ \ee & \ra e^+ e^- (ZZ) \ra 
e^+ e^- + H \non \\ 
(d)  & \ \ {\rm radiation~off~tops} & \ \ \ee & \ra (\gamma,Z) \ra t\bar{t}+H 
\non \end{array} 
\end{eqnarray}

The Higgs--strahlung cross section scales as $1/s$ and therefore dominates at
low energies, while the $WW$ fusion mechanism  has a cross section that rises
like $\log(s/M_H^2)$ and dominates at high energies. The radiative corrections
to these processes are moderate, not exceeding the  few per cent level if the
Fermi constant is used as input. While these corrections have already been 
known for some time for the strahlung process \cite{CRHZ}, they have only
recently been calculated [another ``tour de force"] for the $WW$ fusion process
\cite{CRWW}. At $\sqrt{s} \sim 500$ GeV, the two processes have approximately
the same cross sections, ${\cal O} (100~{\rm fb})$ for the interesting range
100 GeV $\lsim M_H \lsim$ 200 GeV, as shown in Fig.~11.  With an integrated
luminosity $\int {\cal L} \sim 500$ fb$^{-1}$, as expected for instance at the
TESLA machine \cite{Tesla}, approximately 25\, 000 events per year can be
collected in each channel for a Higgs boson with a mass $M_H \sim 150$ GeV.
This sample is more than suffiecient to discover the Higgs boson and to study
its properties in detail. SM--Higgs boson masses of the order of 80\% of the
c.m. energy can be probed, which means that a 800 GeV collider can cover almost
the entire mass range in the SM,  $M_H \lsim 650$ GeV.

\setcounter{figure}{10}
\begin{figure}[hbtp]
\begin{center}
\vspace*{-5.1cm}
\hspace*{-1.2cm}
\epsfig{file=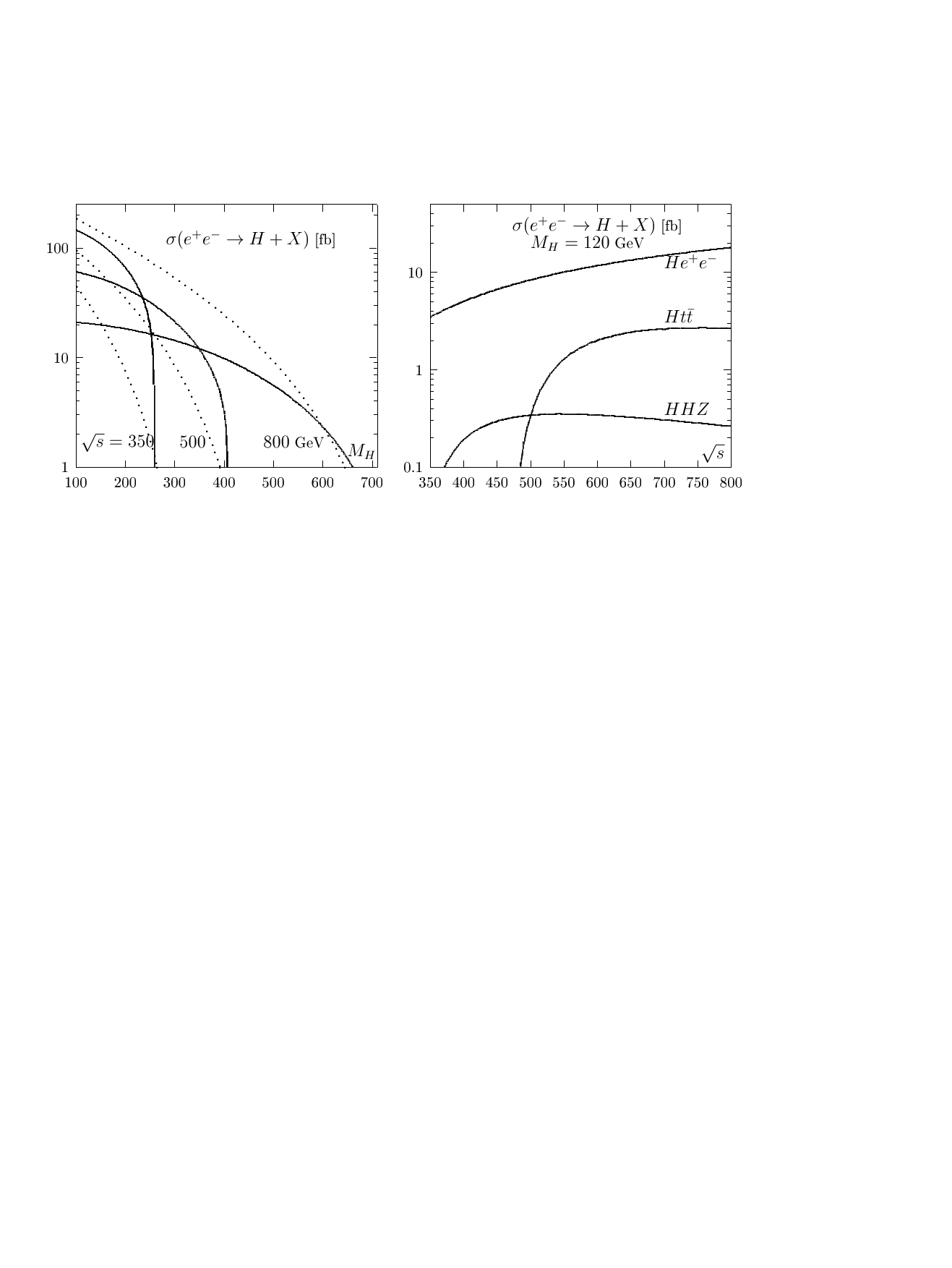,width=20.5cm}
\vspace*{-18.9cm}
\end{center}
\caption[]{Production cross sections of the SM Higgs boson in $e^+ e^-$
collisions in the main processes (left) and in higher order processes (right).}
\vspace*{-0.4cm}
\end{figure}

A stronger case for $\ee$ colliders in the 300--800 GeV energy range is made by
the MSSM. In $\ee$ collisions \cite{E3}, besides the usual strahlung and fusion
processes for $h$ and $H$ production, the neutral Higgs particles can also be
produced pairwise: $\ee \ra A + h/H$.  The cross sections for the strahlung and
pair production as well as those  for the production of $h$ and $H$ are
mutually complementary, coming  with a factor either $\sin^2(\beta- \alpha)$ or
$\cos^2(\beta -\alpha)$. Charged Higgs bosons can be produced pairwise, $\ee
\ra H^+H^-$, through $\gamma,Z$ exchange as well as in top decays for 
$M_{H^\pm} < m_t-m_b$ as at hadron colliders.  

The discussion on the MSSM Higgs production at $\ee$ linear colliders [not
mentioning the $\gamma \gamma$ option of the collider] can be summarized in 
the following points \cite{e+e-,Tesla}: 

$i)$ The Higgs boson $h$ can be detected in the entire range of the MSSM
parameter space, either through the bremsstrahlung process or in pair
production; in fact, this conclusion holds true even at a c.m. energy of 300
GeV and with a luminosity of a few fb$^{-1}$.  

$ii)$ All SUSY Higgs bosons can be discovered at an $\ee$ collider if the $H,A$
and $H^{\pm}$ masses are less than the beam energy; for higher masses, one
simply has to increase $\sqrt{s}$. 

$iii)$ Even if the decay modes of the Higgs bosons are very complicated [e.g.
they decay invisibly], missing mass techniques allow for their detection in
the strahlung process.   

$iv)$ The associated production with  $t\bar{t}$ and $b\bar{b}$ 
states allows for the measurement of the Yukawa couplings and in the $b\bar{b}$ 
case the possible determination of $\tb$ \cite{E4}. $\gamma \gamma \to H,A$  at
photon colliders allows the extension of the mass reach; see also \cite{Jack}
for details.

The determination of the properties of the Higgs bosons can be done in great 
detail in the clean environment of $\ee$ linear colliders \cite{e+e-,Tesla}. 
In the following, relying on analyses done for TESLA \cite{Tesla} [where the 
references for the original studies can be found], we   summarize the possible
measurements in the case of the SM Higgs boson; some of this discussion can 
of course be extended to the the lightest MSSM Higgs particle. 

$\bullet$ The measurement of the recoil $f\bar{f}$ mass in the Higgs--strahlung
process, $\ee \ra ZH\ra H f\bar{f}$ allows a very good determination of the
Higgs boson mass: at $\sqrt{s}=350$ GeV and with a luminosity of $\int {\cal
L}= 500$ fb$^{-1}$, a precision of $ \Delta M_H \sim 50$ MeV can be reached for
$M_H \sim 120$ GeV.  Accuracies  $\Delta M_H \sim 80$ MeV can also be reached
for $M_H=150$ and 180 GeV when the Higgs decays mostly into gauge bosons.   

$\bullet$ The angular distribution of the $Z/H$ in the strahlung process, 
$\sim \sin^2\theta$ at high energy, characterizes the production of a 
$J^P=0^+$ particle. The Higgs spin--parity quantum numbers can also be checked
by looking at correlations in the production $\ee \ra HZ \ra 4f$ or decay $H
\ra WW^* \ra 4f$ processes, as well as in the channel $H \ra \tau^+ \tau^-$ for
$M_H \lsim 140$ GeV. An unambiguous test of the Higgs CP nature  can be made in
the process $\ee \ra t \bar{t}H$ [or at laser photon colliders in the 
loop--induced process $\gamma \gamma \ra H$].  

$\bullet$ The Higgs couplings to $ZZ/WW$ bosons [which are predicted to be
proportional to the masses] can be directly determined by measuring the
production cross sections in the strahlung and the fusion processes.  In
the $\ee \ra H \ell^+ \ell^-$ and $H\nu \bar{\nu}$ processes, the total cross 
section can be measured with a precision less than $\sim$ 3\% at $\sqrt{s}\sim
500$ GeV and with $\int {\cal L}= 500$ fb$^{-1}$. This leads to an accuracy of 
$\lsim$ 1.5\% on the $HVV$ couplings.  

$\bullet$ The measurement of the Higgs branching ratios  is of utmost
importance. For $M_H \lsim 130$ GeV a large variety of ratios can be measured:
the $b\bar{b}, c\bar{c}$  and $\tau^+ \tau^-$ BRs allow us to derive the
relative Higgs--fermion couplings and to check the prediction that they are
proportional to the masses. The gluonic BR is sensitive to the $t\bar{t}H$
Yukawa coupling and to new strongly interacting particles [such as stops in the
MSSM]. The BR into $W$ bosons allows a measurement of the $HWW$ coupling, while
the BR of the loop--induced $\gamma \gamma$ decay is also very important since
it is sensitive to new particles.  

$\bullet$ The Higgs coupling to top quarks, which is the largest coupling in
the SM, is directly accessible in the process where the Higgs boson is radiated
off top quarks, $\ee \ra t\bar{t}H$. For $M_H \lsim 130$ GeV, the Yukawa
coupling  can be measured with a precision of less than 5\% at $\sqrt{s}\sim
800$ GeV with a luminosity $\int {\cal L} \sim 1$ ab$^{-1}$. 

$\bullet$ The total width of the Higgs boson, for masses less than $\sim 200$
GeV, is so small that it cannot be resolved experimentally. However, the
measurement of BR($H \ra WW$) allows an indirect determination of $\Gamma_H$,
since the $HWW$ coupling can be determined from the measurement of the Higgs
cross section in the $WW$ fusion process. [$\Gamma_{\rm tot}$ can
also be derived by measuring the $\gamma \gamma \to H$ cross section at a 
$\gamma\gamma$ collider or BR($H \to \gamma \gamma)$ in $\ee$].

$\bullet$ Finally, the measurement of the trilinear Higgs self--coupling, which
is the first non--trivial test of the Higgs potential, is accessible  in the
double Higgs production processes $\ee \ra ZHH$ [and in the $\ee \ra \nu
\bar{\nu}HH$ process at high energies]. Despite its smallness, the cross 
sections can be determined with an accuracy of the order of 20\% at a 500 GeV 
collider if a high luminosity, $\int {\cal L} \sim 1$ ab$^{-1}$, is available. 

An illustration of the experimental accuracies that can be achieved in the
determination of the mass, CP--nature, total decay width and the various
couplings of the Higgs boson for $M_H=120$ and 140 GeV is shown in Table 1 for
$\sqrt{s}=350$ GeV [for $M_H$ and the CP nature] and $500$ GeV [for
$\Gamma_{\rm tot}$ and all couplings except for $g_{Htt}$] and for $\int {\cal
L}=500$ fb$^{-1}$ [except for $g_{Htt}$ where $\sqrt{s}=1$ TeV and $\int {\cal
L}=1$ ab$^{-1}$ are assumed]. 

\begin{table}[htbp] \vspace*{-6mm} \renewcommand{\arraystretch}{1.8}
\caption{Relative accuracies (in \%) on Higgs boson mass, width and couplings 
obtained at TESLA with $\sqrt{s}=350,500$ GeV and $\int {\cal L}=500$
fb$^{-1}$  (except for top); Ref.~\cite{Tesla}. } \hskip3pc\vbox{\columnwidth=26pc
\begin{tabular}{|c|c|c|c|c|c|c|c|c|c|c|c|}\hline $M_H$ (GeV) & $\Delta M_H$ &
$\Delta {\rm CP}$ & $\Gamma_{\rm tot}$ & $g_{HWW}$ & $g_{HZZ}$ & $g_{Htt}$ &
$g_{Hbb}$ & $g_{Hcc}$ & $g_{H\tau \tau}$ & $g_{HHH}$  \\ \hline $120$ & $\pm
0.033$ & $\pm 3.8$ & $\pm 6.1$ & $\pm 1.2$ & $\pm 1.2$ &  $\pm 3.0$ & $\pm 2.2$
& $\pm 3.7$ & $\pm 3.3$ & $\pm 17$  \\ \hline $140$ & $\pm 0.05$ & $-$ & $\pm
4.5$ & $\pm 2.0$ & $\pm 1.3$ & $\pm 6.1$ & $\pm 2.2$ & $\pm 10$ & $\pm 4.8$ &
$\pm 23$  \\ \hline \end{tabular} } \vspace*{-2mm} \end{table} 

Thus, a high--luminosity $\ee$ linear collider is a very high precision machine
in the context of Higgs physics.  This precision would allow the determination
of the complete profile of the SM Higgs boson, in particular  if its mass is
smaller than $\sim 140$ GeV. It would also allow this particle to be 
distinguished  from the lighter MSSM $h$ boson up to very high values of the
$A$ boson mass, $M_A \sim {\cal O}(1~{\rm TeV})$. This is exemplified in
Fig.~12, where the $(g_{Hbb}, g_{HWW})$ and  $(g_{Hbb}, g_{H \tau \tau})$
contours are shown for $M_H=120$ GeV for a 500 GeV collider with $\int {\cal
L}=500$ fb$^{-1}$. These plots are obtained from a global fit that takes into
account the experimental  correlation between various measurements
\cite{Tesla}.

\begin{figure}[ht!]
\vspace*{-1.3cm}
\begin{center}
\begin{tabular}{c c}
\epsfig{file=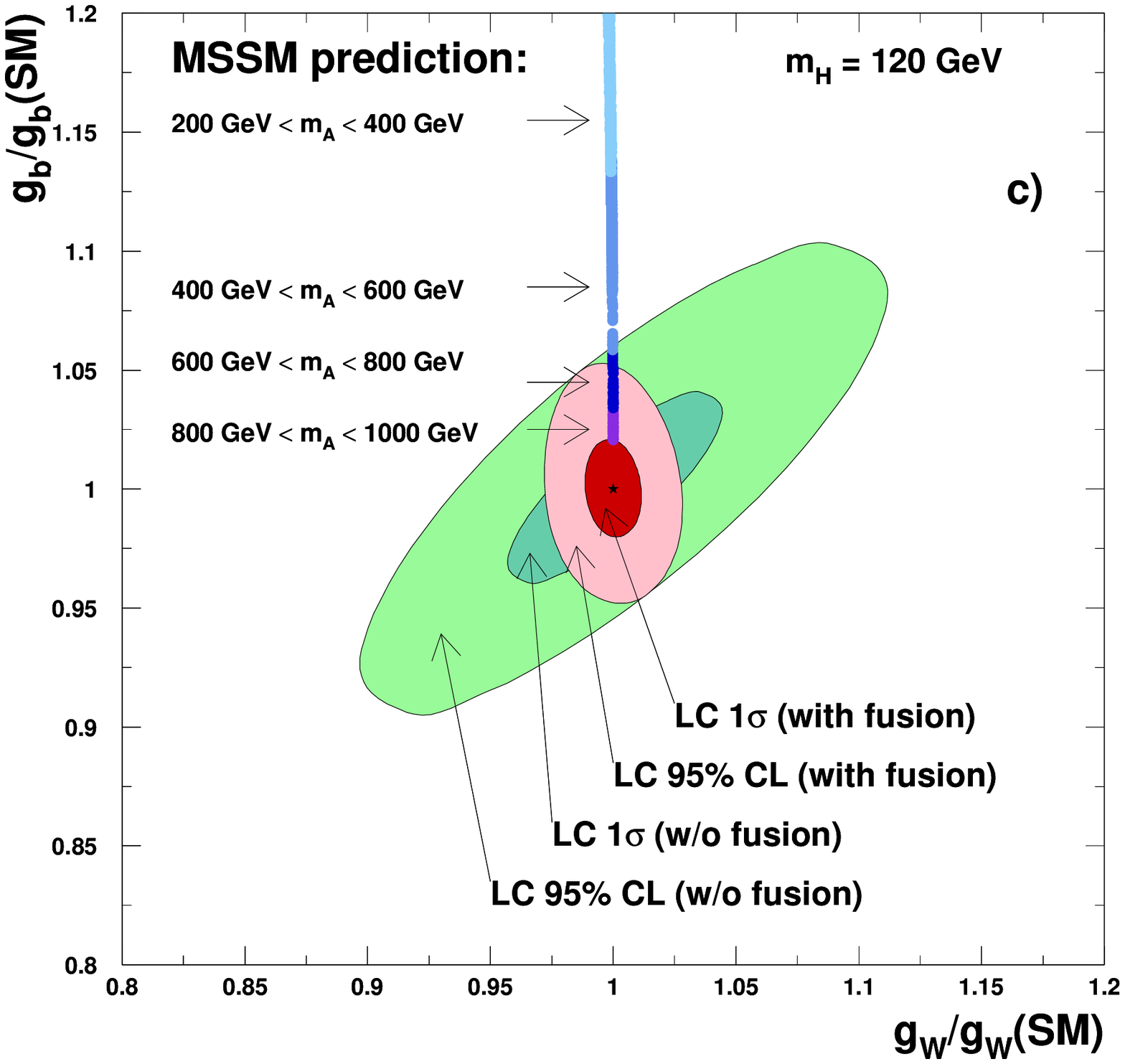,width=0.52\linewidth}\hspace*{-2mm}&\hspace*{-2mm} 
\epsfig{file=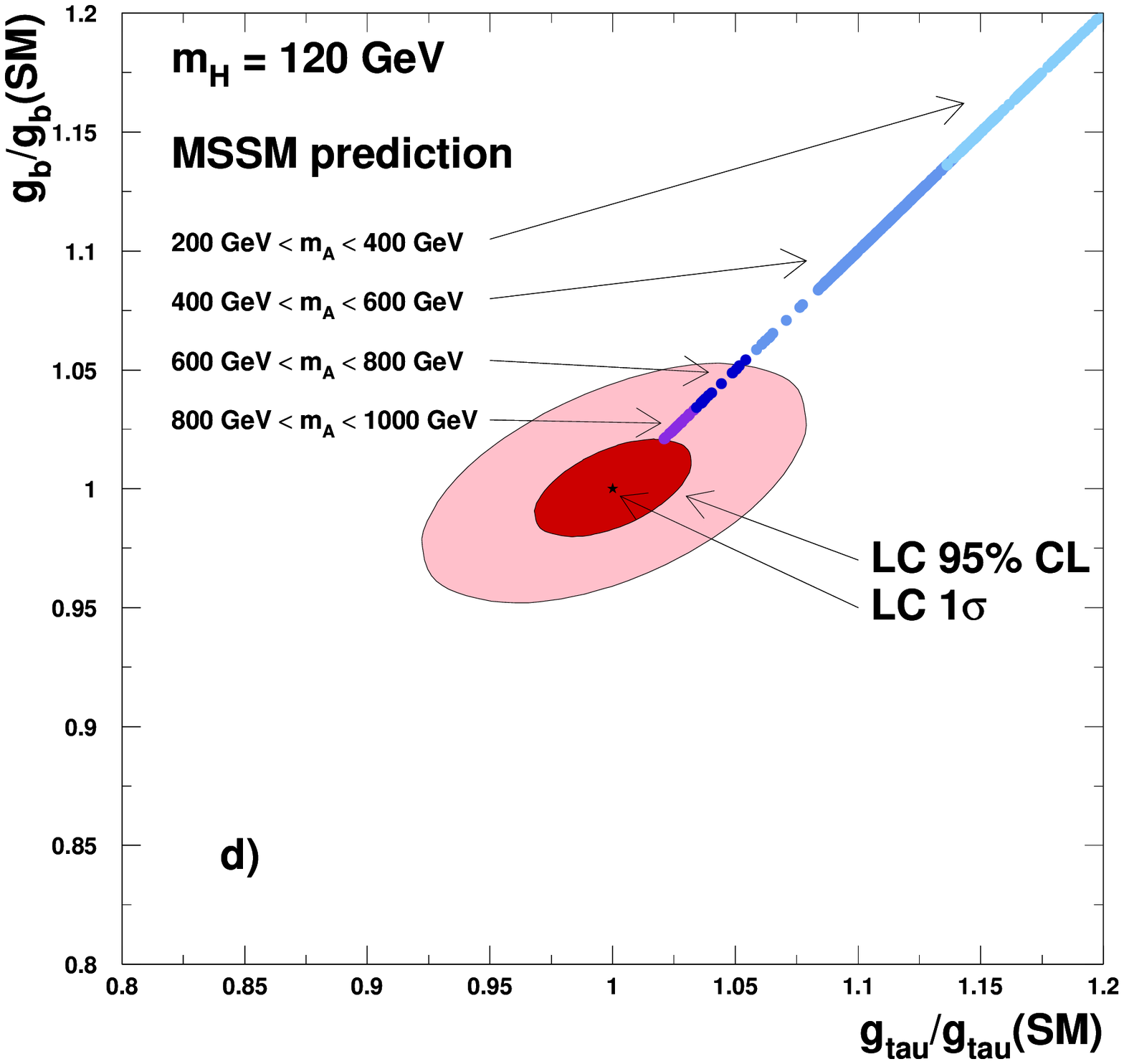,width=0.52\linewidth} \\
\end{tabular}
\caption{Higgs boson coupling determinations at TESLA, for $M_H=120$ GeV
with 500~fb$^{-1}$ of data, and the expected deviations in the MSSM; from
Ref.~\cite{Tesla}.}
\end{center}
\label{fig:hfitter}
\vspace*{-.3cm}
\end{figure}

\subsection*{5. Conclusions} 

In the SM, global fits of the electroweak data favour a light Higgs boson, $M_H
\lsim 200$ GeV;  if the theory is to remain valid up to the GUT scale, the
Higgs boson should be lighter than $200$ GeV. In supersymmetric extensions of
the SM, there is always one light Higgs boson with a mass $M_h \lsim 130$ GeV
in the minimal version [and $M_h \lsim 200$ GeV in more general extensions].
Thus, a Higgs boson is definitely accessible to the next generation of
experiments. The heavier Higgs bosons are expected to have masses in the range
of the electroweak  symmetry breaking scale and can be probed as well.

The detection of a Higgs particle is possible at the upgraded Tevatron for $M_H
\lsim 130$ GeV and is not a problem at the LHC where even much heavier Higgs
bosons can be probed: in the SM up to $M_H \sim 1$ TeV and in the MSSM for
$M_{A,H,H^\pm}$ of order a few hundred GeV, depending on $\tb$. Relatively
light Higgs bosons can also be found at future $\ee$ colliders with
c.m.\,energies $\sqrt{s} \gsim 350$ GeV; the signals are very clear, and the
expected high luminosity allows a thorough investigation of their fundamental
properties.  

In fact, a very important issue once Higgs particles are found, will be to
probe in all its facets the electroweak symmetry breaking mechanism.  In many
aspects, the searches and tests at future $\ee$ colliders are complementary to
those that will be performed at the LHC.  An example can be given in the
context of the MSSM. 

In constrained scenarios, such as the minimal supergravity model, the heavier
$H,A$ and $H^\pm$ bosons tend to have masses of the order of several hundred
GeV and therefore will escape detection at both the LHC and linear collider.
The right--handed panel of Fig.~1 shows the number of Higgs particles in the
$(M_A, \tb$) plane, which can observed at the LHC and in the white area, only
the lightest $h$ boson can be observed.  In this parameter range, the $h$ boson
couplings to fermions and gauge bosons will be almost SM--like and, because of
the relatively poor accuracy of the measurements at the LHC, it would be
difficult to resolve between the SM and MSSM (or extended) scenarios. At $\ee$
colliders such as TESLA, the Higgs couplings can be measured with a great
accuracy, allowing a distinction between the SM and the MSSM Higgs boson to be
made close to the decoupling limit, i.e. for pseudoscalar boson masses, which
are not accessible at the LHC.  This is exemplified in  Fig.~12, where the
accuracy in the determination of the Higgs couplings to $b\bar{b},\tau^+\tau^-$
and $WW$ are displayed, together with the predicted values in the MSSM for
different values of $M_A$.  The two scenarios can be distinguished for
pseudoscalar Higgs masses up to 1 TeV and, thus, beyond the LHC reach.  \bigskip

\nn {\bf Acknowledgements}: I thank the organizers of the Conference, in 
particular S. Banerjee, D.P. Roy and K. Sridhar, for the invitation to the
meeting and for the very nice and warm atmosphere. I would also like to express
all my sympathy to D.P. Roy for the tragedy he has to bear.

\end{document}